\documentclass{iau} 
\usepackage{graphicx}

\title[Model for nulling and moding in radio pulsars] 
{A new model for nulling and moding in radio pulsars}

\author[Jaroslaw Dyks]   
{Jaroslaw Dyks}

\affiliation{Nicolaus Copernicus Astronomical Center, Polish Academy of Sciences, 
87-100 Toru\'n, Poland \\ email: {\tt jinx@ncac.torun.pl}} 

\pubyear{2022}
\volume{363}  
\jname{Neutron Star Astrophysics at the Crossroads: Magnetars and the
Multimessenger Revolution}
\editors{E. Troja \& M. Baring, eds.}
\begin{document}

\maketitle

\begin{abstract}
Single pulse behaviour of radio pulsars is usually interpreted in terms of
the \hbox{{\bf E} $\times$ {\bf B}} drift of a radio beam. It is shown that antisymmetric arrangement of
the radio-bright zones can produce several types of observed single pulse
phenomena: the half-cycle jump in subpulse modulation phase, left-right-middle
subpulse sequence and switching between the core-dominated and
cone-dominated pulsation modes. The geometry can also produce nulling, both
sporadic and intermodal. The model implies that the radio-quiet intervals that
separate the main pulse and interpulse in PSR B0826$-$34  correspond to 
azimuthal breaks in the radio beam, instead of breaks in colatitude.
\end{abstract}

\firstsection 
\section*{The model: beam geometry and simulated pulse stacks}

Single pulse modulation of radio pulsar signals involves the following
striking phenomena: half-cycle jumps in subpulse modulation phase (eg.~PSR B1919+21, 
fig.~3 in Pr\'oszy\'nski \& Wolszczan 1986), left-right-middle sequence of
subpulse appearance within the pulse window (eg.~PSR B1237$+$25, fig.~2 in Hankins \&
Wright 1980) and transitions between cone-dominated and core-dominated
pulsation modes (eg.~PSR B1237$+$25, fig.~5 in Srostlik \& Rankin 2005). A
special form of modulation are nulls that tend to separate different
pulsation modes, but can also appear randomly in the modulation sequence. Changes of profile modes and nulling 
have also been observed in the Galactic centre magnetar PSR J1745$-$2900 (Yan et
al.~2018), however, in this paper I focus on the peculiar single-pulse
modulations as observed in the aforementioned normal pulsars.

\begin{figure}[t]
\begin{minipage}[l]{.75\textwidth}
 \includegraphics[width=\textwidth]{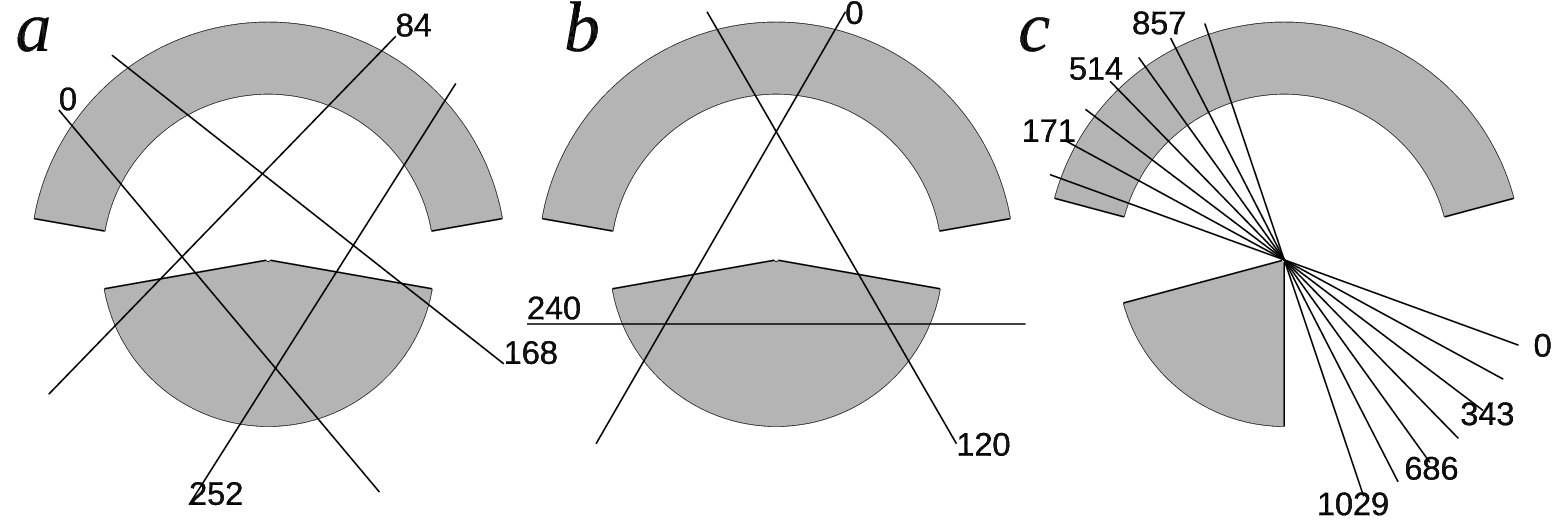}
 \includegraphics[width=\textwidth]{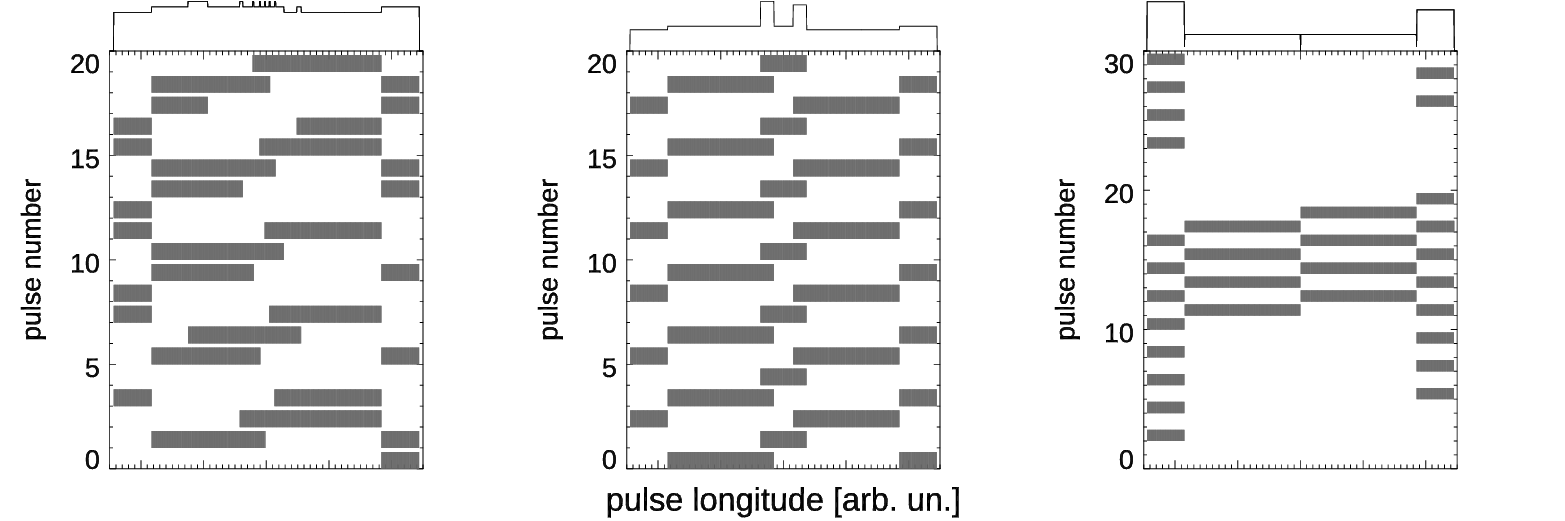}
\end{minipage}%
\hfill
\begin{minipage}[r]{.25\textwidth}
 \includegraphics[width=\textwidth]{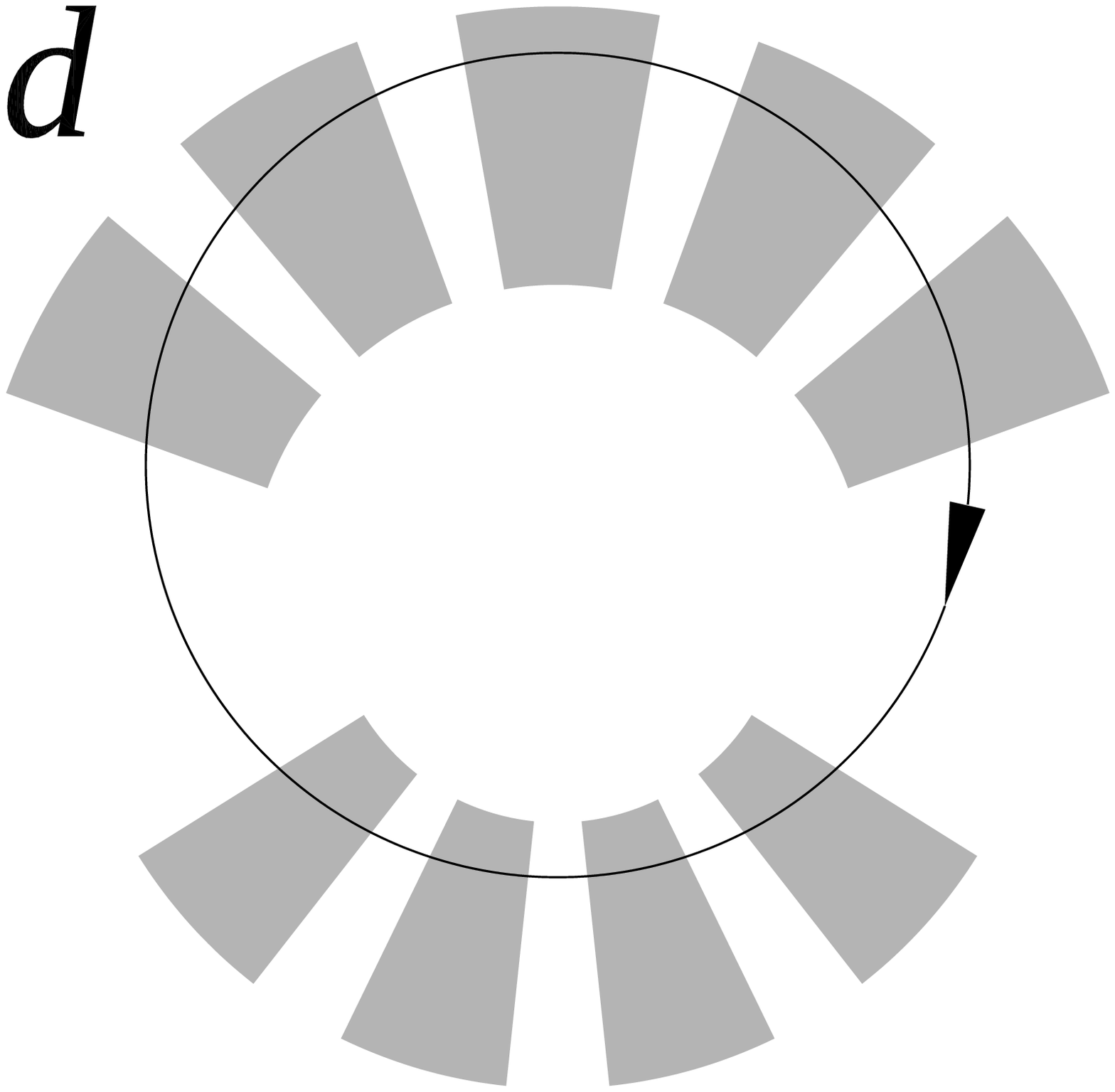}
 \includegraphics[width=\textwidth]{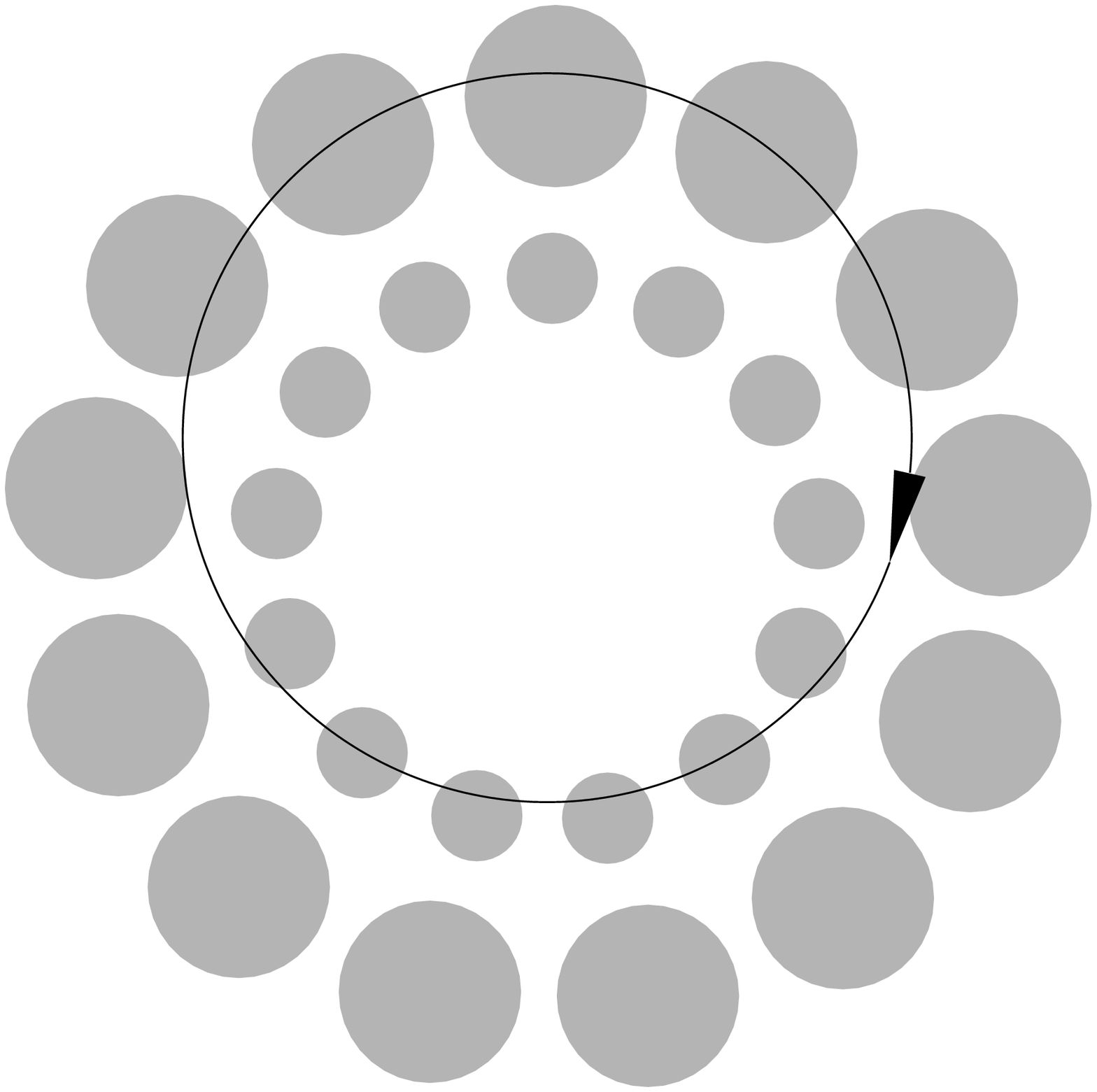}
\vfill
 \label{bgardenbw}
\end{minipage}
\caption{Columns a-c, top: head-on view of a radio beam (grey) with cutting
paths of sightline for $P_d/P=4.3$, $3$ and $2.1$ (left to right). Bottom:
corresponding pulse stacks, with their sum shown on top. Column d: two
beam models for PSR B0826$-$34.  For more details see text.
 }
\label{fione}
\end{figure}

Such types of modulations 
can be explained by an antisymmetrically zonal radio beam which is {\bf E}
$\times$ {\bf B} drifting around the dipole axis at a drift period $P_d$.
The beam is shown in grey in top row of Fig.~\ref{fione} (in columns a, b, and c
which correspond to different drift periods $P_d/P=4.3$, $3$ and $2.1$). 
The solid lines with increasing numbers show consecutive paths
of sightline that cuts the beam at a fixed distance $\beta$ from the beam center once per spin period
$P$ (curvature of the cutting paths is neglected). 
The corresponding modelled pulse stacks are shown in the bottom part of the
figure.

The beam consists of two radio bright zones (one closer to, another further 
from the dipole axis) that are located on the opposite sides of the dipole axis. 
Such antisymmetric arrangement results in the half-cycle jumps in modulation
phase (case a, $P_d=4.3P$). In case b ($P_d=3P$) the left-right-middle sequence 
appears, as explained by the cutting paths on top. When the beam's
inner zone is made asymmetric and $P_d=2.1P$ (case c), then the transitions
between cone-dominated and core-dominated modes appear (column c, bottom). 
When the radio-loud subzones of the beam do not overlap in magnetic azimuth, the 
pulsation modes can be separated by nulls (case c). This happens when the
line of sight is passing through the radio quiet parts of the drifting beam,
while missing the radio loud (grey) parts entirely. Short random nulls are
also possible (case a, pulse no.~5).   

Since the cases b and c are observed in single object (B1237$+$25) 
it seems that both $P_d$ and the beam shape change in this object on
timescale of tens of spin periods. Assuming a near-central cut through the
beam, the nulls observed in 
B1237$+$25 suggest that the beam has non-emitting breaks (voids) within some
intervals of the magnetic azimuth (eg.~cases b,c, top). 
Therefore, a simpler beam with two large voids in azimuth
(column d, top) instead of a void in colatitude (d, bottom)
 can be proposed to explain the low-flux minima between
the main pulse and interpulse in B0826$-$34 (Esamdin et al.~2005). 
The fine azimuthal structure (the split into 4 and 5 subbeams: column d,
top beam) is introduced to reproduce the observation of subpulses in
B0826$-$34 (cf.~fig.~10 in Esamdin et al. 2005), 
however, a time-modulated beam without the fine structure could also
be used. For more
information see Dyks (2021, note, however, the following erratum: the value of $P_d$ for the third case in fig.~4
therein should be corrected to $2.1P$). 

\vskip5mm
\noindent {\bf Acknowledgments} This work was supported by the grant
2017/25/B/ST9/00385 of the National Science Centre, Poland.

\end{document}